\documentclass[conference, 10pt]{IEEEtran}

\usepackage{bbm}
\usepackage{bbold}
\usepackage{amsmath}
\usepackage{amssymb}
\usepackage[dvips]{graphicx}
\usepackage{graphicx}
\usepackage{epsfig}
\usepackage{bigints}
\usepackage{stfloats}
\usepackage{amsthm}
\usepackage{float}
\usepackage{bbold}
\usepackage{bbm, dsfont}
\usepackage{xcolor}

\newtheorem{prop}{Proposition}

\newtheorem{theorem}{Theorem}

\IEEEoverridecommandlockouts

\begin{document}

\title{\vspace{-1ex} Caching in Wireless Small Cell Networks:\\  A Storage-Bandwidth Tradeoff \vspace{-1ex}}  %
\author{
	\IEEEauthorblockN{Syed Tamoor-ul-Hassan, \textit{Student Member, IEEE}, Mehdi Bennis, \textit{Senior Member, IEEE}, \\
	Pedro H. J. Nardelli, \textit{Member, IEEE}, and Matti Latva-aho, \textit{Senior Member, IEEE}} \vspace{-5ex}
	%
	\thanks{This research was supported in part by the TEKES grant 2364/31/2014.
	
	The authors are with Center for Wireless Communications, University of Oulu, Finland. Email: \{tsyed, bennis, nardelli, matti.latva-aho\}@ee.oulu.fi}
}
\providecommand{\keywords}[1]{\textbf{\textit{Index terms---}} #1}
\vspace{-3ex}
\maketitle

\begin{abstract}
Caching contents at the network edge is an efficient mean for offloading traffic, reducing latency and improving users' quality-of-experience. In this letter, we focus on aspects of storage-bandwidth tradeoffs in which small cell base stations are distributed according to a homogeneous Poisson point process and cache contents according to a given content popularity distribution, subject to storage constraints. We provide a closed-form expression of the cache-miss probability, defined as the probability of not satisfying users’ requests over a given coverage area, as a function of signal-to-interference ratio, cache size, base stations density and content popularity.
In particular, it is shown that for a given minimum cache size, the popularity based caching strategy achieves lower outage probability for a given base station density compared to uniform caching. Furthermore, we show that popularity based caching attains better performance in terms of cache-miss probability for the same amount of spectrum.
\end{abstract}
\vspace{-0.1cm}
\section{Introduction}
\label{sec:Intro}
Heterogeneous wireless networks, comprising of small cell base stations (SBSs) underlying the macrocellular network, is a promising solution to improve capacity, coverage and users' quality-of-experience (QoE) \cite{Ref3:SCN_2}.
At the same time, the unplanned deployment of dense small cells further renders these problems very challenging \cite{ref:Spec_Utilization1} \cite{ref:Spec_Utilization2}. Recently, content caching at the network edge (i.e., SBSs, user devices etc.) has been proposed as a cost-efficient solution to offload cellular/backhaul traffic, while satisfying users' QoE and alleviating network congestion \cite{Ref:Boc_5G} \cite{Ref6:Intro1}.  Although content caching and spectrum sharing received significant attention in recent years, most of the existing works consider spectrum allocation and caching separately, and little work has been done in exploring their joint benefits. In this respect, \cite{Ref7:lit_review3} proposes a collaborative caching mechanism for wireless multimedia streaming exploiting storage and bandwidth auctions. However, the distribution of caches is considered fixed.  Meanwhile, \cite{Ref7:lit_review4} studies the gains of caching in a stochastically distributed small cell scenario, where the effects of storage size and file popularity are investigated. Furthermore, the work in \cite{Ref7:lit_review5} characterizes the outage probability in small cell networks for uniform content distribution. However, the impact of spectrum has not been addressed in \cite{Ref7:lit_review4} \cite{Ref7:lit_review5}. \\
\indent Unlike \cite{Ref7:lit_review5}, this letter characterizes the outage probability in serving user requests over a coverage area by jointly exploiting spectrum and caching. By considering the distribution of macro cellular base stations (MBSs) (with large cache) and SBSs (with limited cache) as Poisson point processes (PPPs), we derive a closed form expression of the outage probability for a given content distribution. The outage probability is defined as the probability of not satisfying users requests over a given coverage area, as a function of signal-to-interference ratio (SIR), cache size and SBSs density. We emphasize that the proposed model is pessimistic in the sense that it provides a lower bound on the overall network performance and focuses on the regime with large SBS density. For spectrum allocation, we define the spectrum access factor as a measure of spectrum accessed by the SBSs. 
We also show the interplay of caching and spectrum sharing in such heterogeneous networks, drawing insights on the tradeoff between cache size and spectrum access factor, for different content popularity models.
\vspace{-0.25cm}
\section{System Model}
\begin{figure}[t]%
\centering
\includegraphics[scale = 0.35]{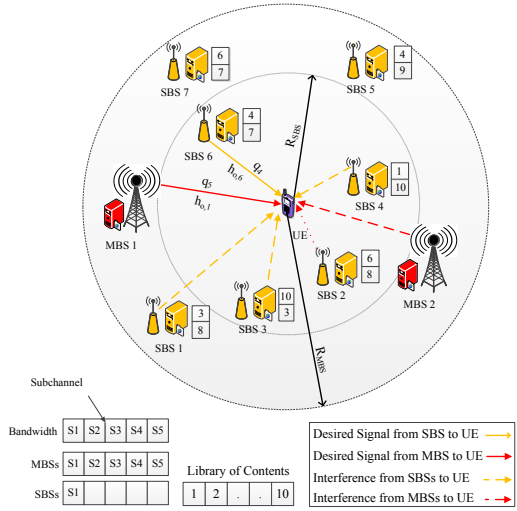}
\caption{The user requests content $q_4$ and $q_5$ cached by SBS 6 and MBS 1 within the threshold distance $R_{{\mathrm{SBS}}}$ and $R_{{\mathrm{MBS}}}$ respectively. SBSs access 1 sub-channel out of 5 sub-channels, leading to a spectrum access factor of 0.2.}
\vspace{-0.1cm}
\label{fig:Net_Diagram5}
\end{figure}

Consider the downlink transmission of a two-tier cellular network comprising of SBSs, underlying MBSs in a two-dimensional Euclidean plane $\mathbb{R}^2$ as shown in Fig. \ref{fig:Net_Diagram5}. Hereafter, we will use the subscripts ``$\mathrm{SBS}$'' and ``$\mathrm{MBS}$'' to refer to the system variables associated to the small- and macro-base-stations, respectively. We assume that MBSs are deployed according to a homogeneous PPP, $\Phi_{\mathrm{MBS}} = \{X_i\}$ with intensity $\lambda_{\mathrm{MBS}}$ (MBSs per square meter), where $X_i$ represents the spatial location of the $i$-th MBS (in relation to the origin). Similarly, SBSs distribution is modeled as a homogeneous PPP, $\Phi_{\mathrm{SBS}} = \{Y_j\}$, with intensity $\lambda_{\mathrm{SBS}}$ (SBSs per square meter), such that $\lambda_{\mathrm{SBS}} \gg \lambda_{\mathrm{MBS}}$ and $Y_j$ denotes the spatial location of the $j$-th SBS. We consider a spectrum configuration where the bandwidth $W$ is divided into $B$ equal sub-channels and each SBS randomly (and independently) accesses $\beta B$ sub-channels, such that $\beta \in [0,1]$. Hereafter, $\beta$ is referred to as spectrum access factor. The transmit power of MBS and SBS is represented by $p_{\mathrm{MBS}} = \frac {p_{\mathrm{maxMBS}}} {B}$ and $p_{\mathrm{SBS}} = \frac {p_{\mathrm{maxSBS}}} {\beta B}$, respectively where $p_{\mathrm{maxMBS}}$ and $p_{\mathrm{maxSBS}}$ is the maximum transmit power of MBS and SBS, respectively. \\
\indent Let each MBS be equipped with a large storage to cache contents from a given library $\mathcal{C} = \{1,2, ..., |\mathcal{C}|\}$, with size $|\mathcal{C}|$. All library contents are of equal size (in bits). Each SBS, in turn, is equipped with a cache storage of size $d$ and stores a subset of library contents. For simplicity, we express the cache size in terms of normalized cache size where $\tilde{d} = d/|\mathcal{C}|$.  
Let $\Omega_{j} \subseteq \mathcal{C}$ represent the specific content(s) cached at the $j$-th SBS. We assume that SBSs cache contents following two different policies: a uniform caching policy (UCP), where the SBSs randomly caches contents regardless of their popularity, or popularity-based caching policy (PCP), where the SBSs hold the $d$ most popular contents. \\
\indent We assume that a reference user -- a mobile user -- is located at the origin $o$ and analyze the system performance for the reference link assuming it requests a given content $c$ independently at each time slot. We analyze two different distribution of users' requests: (i) uniform, where the contents are equally popular, and (ii) Zipf where the contents' popularity follows a power-law distribution \cite[Sec.~IV-C]{Ref8:Caching_Ref2}. \\
\indent We define a threshold distance $R_{\mathrm{MBS}}$ and $R_{\mathrm{SBS}}$ that specifies the maximum distance over which the MBS and SBS serves the user's content request respectively.
The reference user is associated to the nearest SBS with the cached content within $R_{\mathrm{SBS}}$.
If no SBS has cached the requested content, MBSs may fulfill the user request. In this scenario, the user attempts to associate to the nearest MBS within $R_{\mathrm{MBS}}$ such that $R_{\mathrm{MBS}} > R_{\mathrm{SBS}}$. 
At each transmission, we assume a standard power loss propagation model with path loss exponent $\alpha > 2$. The multi-path effects are modeled as Rayleigh fading whose gains $h_{o}$ are exponentially distributed with mean $h_{o} \sim \exp(1)$. Furthermore, we assume an interference-limited regime and therefore, neglect the effects of Additive White Gaussian Noise (AWGN). Then, the SIR is computed as:
\vspace{-0.1cm}
\begin{equation}
\mathrm{SIR}_{o,z} = \frac {p_z h_{o,z} R_{z}^{-\alpha}} { I_{o,z}}, \ \ z \in \{\mathrm{SBS}, \mathrm{MBS}\},
\vspace{-0.1cm}
\end{equation}
where $p_z$ is the transmit power of $z$, 
$R_z$ is the (random) distance of the reference link and $I_{o,z}$ is the interference experienced by the reference user. The interference $I_{o,z}$ at the reference user is the interference from all the MBSs and SBSs transmitting on the same frequency. As the interference from SBSs is a thinned PPP, $\tilde{\Phi}_{\mathrm{SBS}}$ with density $\beta \lambda_{\mathrm{SBS}}$, then the interference experienced by the reference user served by an SBS and MBS respectively is given by:
\vspace{-0.4cm}

\footnotesize
\begin{flalign}
\label{eq:Interference}
& I_{o, \mathrm{SBS}} = \sum\limits_{X_i \in \Phi_{\mathrm{MBS}}} p_{\mathrm{MBS}} h_{o,\mathrm{MBS}} X_i^{-\alpha} + \sum\limits_{Y_j \in \tilde{\Phi}_{\mathrm{SBS}} \setminus \{o\}} p_{\mathrm{SBS}} h_{o,\mathrm{SBS}} Y_j^{-\alpha}, & \\
\vspace{-0.2cm}
& I_{o, \mathrm{MBS}} = \sum\limits_{X_i \in \Phi_{\mathrm{MBS}} \setminus \{o\}} p_{\mathrm{MBS}} h_{o,\mathrm{MBS}} X_i^{-\alpha} + \sum\limits_{Y_j \in \tilde{\Phi}_{\mathrm{SBS}}} p_{\mathrm{SBS}} h_{o,\mathrm{SBS}} Y_j^{-\alpha}. 
\end{flalign}
\normalsize
\vspace{-0.7cm}
\section{Analytic Results}
In this section, we  derive a closed-form expression of the outage probability for  serving user requests.
The analysis considers the content distribution in a static network where a single network realization considers $\Phi_{\mathrm{MBS}}$, $\Phi_{\mathrm{SBS}}$ and $\Omega_{j}$ independently.
\vspace{-0.1cm}
\subsection{Cache Hit Probability}
\vspace{-0.1cm}
\label{ssec:Cache_Hit_Probability}
In order to define the cache hit probability, we first introduce a content replication $P_c$ defined as the probability that content $c$ is cached (replicated) at the SBS, expressed as: 
\vspace{-0.2cm}
\begin{equation}
P_c = \mathbb{P} \bigg(c \in \Omega_{j}\bigg),
\vspace{-0.1cm}
\end{equation}
\vspace{-0.1cm}
where $c$ is ordered by the content popularity (i.e. $c=1$ is the most popular content, $c=2$ the second most popular and so on). As mentioned earlier, SBSs employ two replication strategies, namely UCP and PCP.
For UCP, $P_c$ is the fraction of library contents stored by SBS (i.e., $P_c = d/|\mathcal{C}|$). For PCP, the SBSs cache $d$ most popular contents. In this case, $P_c$ is given by:
\vspace{-0.2cm}
\begin{equation}
P_c = \begin{cases} 1 & \quad c \leq d  \\
                    0 & \quad \text{otherwise}
	     \end{cases}.
\vspace{-0.1cm}
\end{equation}
\indent User requests, represented by $q_c$, are modeled as uniform or Zipf distribution. While in the former user randomly requests library contents, the latter considers different file popularity. For the Zipf distribution, content requests are ranked from the most popular to the least such that the request for a content $c \in \mathcal{C}$ is:
\vspace{-0.3cm}
\begin{equation}
q_c = \frac {1} {c^{\delta}} \bigg(\sum_{i=1}^{|\mathcal{C}|} \frac {1} {i^{\delta}}\bigg)^{-1}.
\vspace{-0.2cm}
\end{equation}
\noindent Note that $\delta$ models the skewness of the popularity profile. For $\delta=0$ the popularity profile is uniform over files, and becomes more skewed as $\delta$ grows larger \footnote{The value of $\delta$ is 0.8 and 1.2 for user-generated contents and video-on-demand contents respectively \cite{Ref8:Caching_Ref2}.}. \\
\indent Based on content replication, the cache hit probability is defined as the probability of existence of the content within $R_{\mathrm{SBS}}$ or $R_{\mathrm{MBS}}$. Therefore, for a given reference user, the existence of the closest SBS within $R_{\mathrm{SBS}}$ is given by:
\vspace{-0.3cm}
\begin{align}
\label{eq:CacheHit_SBS}
\mathbb{P}_{\mathrm{SBS}} (c) & = 1 - \mathbb{P}( \nexists \textup{ SBS with } c \textup{ within } R_{\mathrm{SBS}}) & \nonumber \\
& \ = 1 - e^{-\beta B \lambda_{\mathrm{SBS}} P_c \pi (R_{\mathrm{SBS}})^2}.
\vspace{-0.7cm}
\end{align}

\vspace{-0.35cm}
\indent For the MBSs, we proceed in a similar way, but considering that they have all contents in their library. Then,
\vspace{-0.3cm}
\begin{flalign}
\mathbb{P}_{\mathrm{MBS}} & = 1 -\mathbb{P}( \nexists \textup{ MBS }  \textup{ within } R_{\mathrm{MBS}} ) \nonumber \\
& = 1 - e^{-\lambda_{\mathrm{MBS}} \pi (R_{\mathrm{MBS}})^2}.
\vspace{-0.7cm}
\end{flalign}

\vspace{-0.35cm}
\indent By considering the fact that the reference user always associates to the closest SBS having the requested content, the probability density function $f_{c,\mathrm{SBS}}(r)$ of the distance $R=r$, given that the content $c$ exists within $R_{\mathrm{SBS}}$, is \cite{Ref8:Caching_Ref4}:
\small
\vspace{-0.2cm}
\begin{equation}
\label{eq:content_dist}
f_{c,\mathrm{SBS}}(r) = 2 \pi \beta B \lambda_{\mathrm{SBS}} P_c r \frac {e^{- \beta B \lambda_{\mathrm{SBS}} P_c \pi r^2}} {1 - e^{-\beta B \lambda_{\mathrm{SBS}} P_c \pi (R_{\mathrm{SBS}})^2}}, \ \ \textcolor[rgb]{0,0,1}{\beta P_c > 0}.
\vspace{-0.1cm}
\end{equation}
\vspace{-0.2cm}
\normalsize
For MBS, the probability density function $f_{c, \mathrm{MBS}}(r)$ is:
\begin{equation}
\label{eq:content_dist1}
f_{c,\mathrm{MBS}}(r) = 2 \pi \lambda_{\mathrm{MBS}} r \frac {e^{- \lambda_{\mathrm{MBS}} \pi r^2}} {1 - e^{-\lambda_{\mathrm{MBS}} \pi (R_{\mathrm{MBS}})^2}}.
\end{equation}

\subsection{Outage Probability}
Let $\mathcal{T}_{\mathrm{SBS}}(c)$ be the event that the requested content $c$ has been transmitted successfully given that $c$ has been cached by a SBS within $R_{\mathrm{SBS}}$. Then, the outage probability of transmitting content $c$ over $R_{\mathrm{SBS}}$, denoted by $\mathbb{P}_{\mathrm{out},\mathrm{SBS}}(c)$ is:
\vspace{-0.2cm}
\begin{equation}
\mathbb{P}_{\mathrm{out},\mathrm{SBS}}(c) = 1 - \mathbb{P}[\mathcal{T}_{\mathrm{SBS}}(c)],
\end{equation}
where, for a given SIR threshold $\gamma$, $\mathbb{P}[\mathcal{T}_{\mathrm{SBS}}(c)] = \mathbb{P}(\mathrm{SIR}_{o,\mathrm{SBS}} (c)>\gamma)$. Note that, due to the association procedure, the SIR implicitly depends on the content requested. \\
\indent For the MBS, all the contents are cached and therefore the outage probability is the same for all $c \in \mathcal{C}$, given that there exists at least one MBS within $R_{\mathrm{MBS}}$. In this case,
\vspace{-0.2cm}
\begin{equation}
\mathbb{P}[\mathcal{T}_{\mathrm{MBS}}(c)] = \mathbb{P}(\mathrm{SIR}_{o,\mathrm{MBS}}>\gamma).
\vspace{-0.1cm}
\end{equation}
\indent The probability that transmission is successful, conditioned on the random variable $R$, representing the distance between the reference user and the base-station associated with it, is:
\vspace{-0.2cm}

\footnotesize
\begin{flalign}
\mathbb{P}(\mathrm{SIR}_{o} > \gamma \ |  \ R ) & = \mathbb{P}\bigg[\frac {p h_{o} R^{-\alpha}} {I_o} > \gamma \ | \ R \bigg] 
= \mathbb{P} \bigg[h_{o} > \frac {\gamma I_o R^{\alpha}} {p} \ | \ R \bigg] & \nonumber \\
\ \ \ \ \ \ \ \ \ \ =& \mathbb{E}_{I_o} \bigg[\mathbb{P} \bigg[h_{o} > \frac {\gamma I_o R^{\alpha}} {p} \ | \ R, I_o \bigg] \bigg] & \nonumber \\
=& \mathbb{E}_{I_o | R} \bigg[\exp\bigg(\frac {- \gamma I_o R^{\alpha}} {p} \bigg) \bigg] = \mathcal{L}_{I_o | R} \bigg(\frac {\gamma R^{\alpha}} {p} \bigg).
\end{flalign}
\normalsize
\begin{prop}
The Laplace transform of the interference experienced by the reference user served by an SBS or MBS over the random distance $R_1$ and $R_2$ respectively, defined in \eqref{eq:Interference}, is given by:
\vspace{-0.3cm}

\footnotesize
\begin{flalign}
\label{eq:Laplace_Transform}
\mathcal{L}_{I_{o, \mathrm{SBS}}} \bigg(\frac {\gamma R_{1}^{\alpha}} {p_{\mathrm{SBS}}} \bigg) = \exp(- \pi R_{1}^2 (\lambda_{\mathrm{MBS}} k_1 + \beta \lambda_{\mathrm{SBS}} k_2)), & \\
\mathcal{L}_{I_{o, \mathrm{MBS}}} \bigg(\frac {\gamma R_{2}^{\alpha}} {p_{\mathrm{MBS}}} \bigg) = \exp(- \pi R_{2}^2 (\lambda_{\mathrm{MBS}} k_3 + \beta \lambda_{\mathrm{SBS}} k_4)). &
\end{flalign}
\normalsize
where $k_1$ and $k_2$ are given by:
\footnotesize
\begin{equation}
\begin{split}
k_1 &= \bigg(\frac {\gamma p_{\mathrm{MBS}}} {p_{\mathrm{SBS}}}\bigg)^{2/\alpha} \int_{ \big(\frac {p_{\mathrm{SBS}}} {\gamma p_{\mathrm{MBS}}}\big)^{2/\alpha}}^{\infty} \frac {1} {1 + u_1^{\alpha/2}} \mathrm{d}u_1, \nonumber \\
k_2 &= k_3 = \gamma^{2/\alpha} \int_{\gamma^{-2/\alpha}}^{\infty} \frac {1} {1 + u_2^{\alpha/2}} \mathrm{d}u_2, \nonumber \\
k_4 &= \bigg(\frac {\gamma p_{\mathrm{SBS}}} {p_{\mathrm{MBS}}}\bigg)^{2/\alpha} \int_{ \big(\frac {p_{\mathrm{MBS}}} {\gamma p_{\mathrm{SBS}}}\big)^{2/\alpha}}^{\infty} \frac {1} {1 + u_4^{\alpha/2}} \mathrm{d}u_4. \nonumber
\end{split}
\end{equation}
\normalsize
\end{prop}
\vspace{-0.2cm}
\begin{IEEEproof}
Considering the conditional Laplace transform;
\small
\begin{flalign}
\label{eq-int-1}
\mathcal{L}_{I_{o,{\mathrm{SBS}}}} (t) & = \mathbb{E}_{I_{o,{\mathrm{SBS}}}} (e^{-t I_{o,\mathrm{SBS}}}), & \nonumber \\
\overset{(a)}{{=}} & \ \mathbb{E}_{\Phi_{\mathrm{MBS}}, h_{o,\mathrm{MBS}}} \bigg[\exp(-t \sum_{i \in \Phi_{\mathrm{MBS}}} p_{\mathrm{MBS}} h_{o,\mathrm{MBS}} X_i^{-\alpha}) \bigg]  \times & \nonumber \\
& \mathbb{E}_{\tilde{\Phi}_{\mathrm{SBS}}, h_{o,\mathrm{SBS}}} \bigg[\exp(-t \sum_{j \in \tilde{\Phi}_{\mathrm{SBS}} \setminus \{o\}} p_{\mathrm{SBS}} h_{o,\mathrm{SBS}} Y_j^{-\alpha}) \bigg],  &
\end{flalign}
\normalsize
\vspace{-0.4cm}

where $t = \frac {\gamma R^{\alpha}} {p_{\mathrm{SBS}}}$ and $(a)$ is the definition of Laplace transform. By using the definition of probability generating functional of PPP, i.i.d of $\Phi_{\mathrm{SBS}}$ and $\Phi_{\mathrm{MBS}}$ and exponential interference distribution, the simplification yields \eqref{eq:Laplace_Transform} \cite{Ref9:Coverage}. The laplace tranform of interference experienced by a reference user served by an MBS follows the same procedure as above.
\end{IEEEproof}
\begin{prop}
\label{prop:PoutSBS}
Given the SIR threshold $\gamma$, the outage probability of content $c$ requested by reference user over $R_{\mathrm{SBS}}$ is:
\vspace{-0.2cm}
\footnotesize
\begin{equation}
\label{eq:Out_Rs}
\vspace{-0.1cm}
\mathbb{P}_{\mathrm{out},\mathrm{SBS}}(c) = 1 - \frac {\beta B P_c \lambda_{\mathrm{SBS}} \bigg[1 - e^{-\pi R_{\mathrm{SBS}}^2 \big(k_1 \lambda_{\mathrm{MBS}} + (k_2 + P_c B) \beta \lambda_{\mathrm{SBS}} \big) }\bigg]} {\big(k_1 \lambda_{\mathrm{MBS}} + (k_2 + P_c B) \beta \lambda_{\mathrm{SBS}} \big) (1 - e^{-\beta B \lambda_{\mathrm{SBS}} P_c \pi R_{\mathrm{SBS}}^2})},
\end{equation}
\normalsize
\end{prop}
\begin{IEEEproof}
The outage probability in serving the user request for $c$ over the distance $R$, whose pdf is given in \eqref{eq:content_dist}, is:
\small
\begin{flalign}
\ \ \ \ \ \ \ \ \ \ \mathbb{P}_{\mathrm{out}, \mathrm{SBS}}(c) &= \mathbb{E}_R[\mathbb{P}(\mathrm{SIR}_{o,{\mathrm{SBS}}} (c) < \gamma) \ | \ R = r] & \nonumber \\
&= \int_{0}^{R_{\mathrm{SBS}}} \bigg[ 1 - \mathcal{L}_{I_{o,{\mathrm{SBS}}}} \bigg(\frac {\gamma r^{\alpha}} {p_{\mathrm{SBS}}} \bigg) \bigg] f_{c,\mathrm{SBS}}(r) \mathrm{d}r. & \nonumber
\end{flalign}
\normalsize
Using \eqref{eq:content_dist} to solve the above integral, we find \eqref{eq:Out_Rs}.
\end{IEEEproof}
\begin{prop}
The outage probability of content $c$ requested by the reference user over a threshold distance $R_{\mathrm{MBS}}$, is:
\footnotesize
\begin{equation}
\label{eq:out_Rm}
\mathbb{P}_{\mathrm{out},\mathrm{MBS}}(c) = 1 - \frac { \lambda_{\mathrm{MBS}} \bigg[1 - e^{-\pi R_{\mathrm{MBS}}^2 \left( \lambda_{\mathrm{MBS}} (k_3 + 1) + \beta \lambda_{\mathrm{SBS}} k_4 \right) }\bigg] } {\left( \lambda_{\mathrm{MBS}} (k_3 + 1) + \beta \lambda_{\mathrm{SBS}} k_4 \right) \big(1 - e^{-\lambda_{\mathrm{MBS}} \pi R_{\mathrm{MBS}}^2} \big)},
\end{equation}
\end{prop}
\normalsize
This proposition's proof follows the same steps as before.
\begin{theorem}
\label{Coro_1}
For a given $P_c$, $R_{\mathrm{SBS}}$ and $R_{\mathrm{MBS}}$, the outage probability in serving a user requesting content $c$ is given by:
\vspace{-0.5cm}
\begin{flalign}
\mathbb{P}_{\mathrm{out}}(c) = & \mathbb{P}_{\mathrm{SBS}}(c) \mathbb{P}_{\mathrm{out},\mathrm{SBS}}(c) + (1 - \mathbb{P}_{\mathrm{SBS}}(c)) \times \bigg[\mathbb{P}_{\mathrm{MBS}} & \nonumber \\
& \mathbb{P}_{\mathrm{out},\mathrm{MBS}}(c) + (1 - \mathbb{P}_{\mathrm{MBS}}) \bigg].
\end{flalign}
\end{theorem}
\normalsize
\begin{IEEEproof}
Theorem \ref{Coro_1} imply that if the content has been cached by a SBS within $R_{\mathrm{SBS}}$, the outage probability of serving the user request is given by \eqref{eq:Out_Rs}. Otherwise, the MBS within $R_{\mathrm{MBS}}$ serves the reference user with the desired content, given by \eqref{eq:out_Rm}. Therefore, the mean outage probability is the linear combination of \eqref{eq:Out_Rs} and \eqref{eq:out_Rm}.
\end{IEEEproof}
Meanwhile, the average outage probability is given by:
\begin{equation}
\mathbb{P}_{\mathrm{out},\mathrm{avg}} = \sum_{c \in \mathcal{C}} q_c \mathbb{P}_{\mathrm{out}}(c).
\end{equation}
\vspace{-0.5cm}
\section{Numerical Results}
\begin{figure}[t]%
\centering
\vspace{0.1cm}
\includegraphics[scale = 0.25]{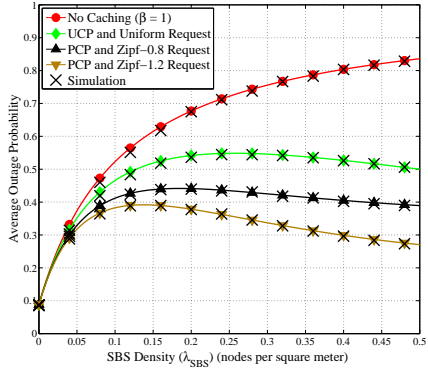}
\vspace{-0.5cm}
\caption{Outage probability vs. SBS density. $\lambda_{\mathrm{MBS}} = 0.0001\mathrm{m}^{-2}$, $R_{\mathrm{SBS}} = 5$m, $R_{\mathrm{MBS}} = 250$m, $\gamma = -10$dB, $\alpha = 4$, $\tilde{d} = 0.3$, $\beta = 0.05$.}
\vspace{-0.15cm}
\label{fig:SBS_Tradeoff}
\end{figure}

\begin{figure}[t]%
\centering
\includegraphics[scale = 0.255]{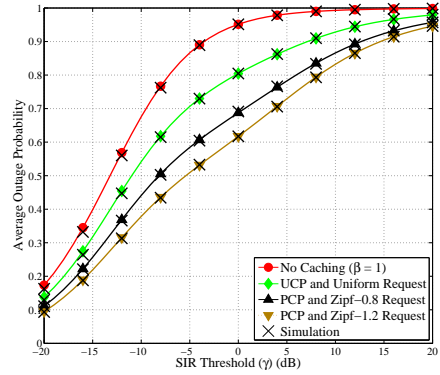}
\vspace{-0.5cm}
\caption{Outage probability vs SIR threshold. $\lambda_{\mathrm{SBS}} = 0.2\mathrm{m}^{-2}$, $\lambda_{\mathrm{MBS}} = 0.0001\mathrm{m}^{-2}$, $R_{\mathrm{SBS}} = 5$m, $R_{\mathrm{MBS}} = 250$m, $\alpha = 4$, $\tilde{d} = 0.3$, $\beta = 0.05$.}
\vspace{-0.15cm}
\label{fig:Threshold_Tradeoff}
\end{figure}

We present here a numerical analysis of the outage probability, validating our analytical results with Monte-Carlo simulations for a network grid of size $1000\mathrm{m} \times 1000\mathrm{m}$. We generate $10,000$ requests from a library of size $100$ according to the request distribution described in Sec. \ref{ssec:Cache_Hit_Probability}. For each request, the outage probability is evaluated over $2000$ network realizations. In addition, we assume $B=1$ while the maximum transmit powers of MBS and SBS are set to 43dbm and 23dbm, respectively.   
Fig. \ref{fig:SBS_Tradeoff} shows the outage probability for various caching strategies as a function of SBS density by considering uniform and Zipf content request distributions. It can be clearly seen that the analytical results match very well the simulation results, thus validating the analytical expression. Without caching, increasing $\lambda_{\mathrm{SBS}}$ only increases interference, resulting in an increased outage probability. When SBSs cache contents, the outage probability varies with increasing $\lambda_{\mathrm{SBS}}$ depending on the content distribution. For each of the content distribution, the outage probability first increases before levelling off. Under Zipf distribution, the outage probability maximizes for smaller $\lambda_{\mathrm{SBS}}$ as SBSs only cache the most popular contents. Under uniform distribution, the outage probability is maximized for a relatively larger $\lambda_{\mathrm{SBS}}$, thus requiring more SBSs to fulfill user requests. In addition, the cache hit probability given by (7) is an increasing function of $\lambda_{\mathrm{SBS}}$. \\
\begin{figure}[b]%
\centering
\vspace{-0.3cm}				
\includegraphics[height = 6.50cm, width = 7.25cm]{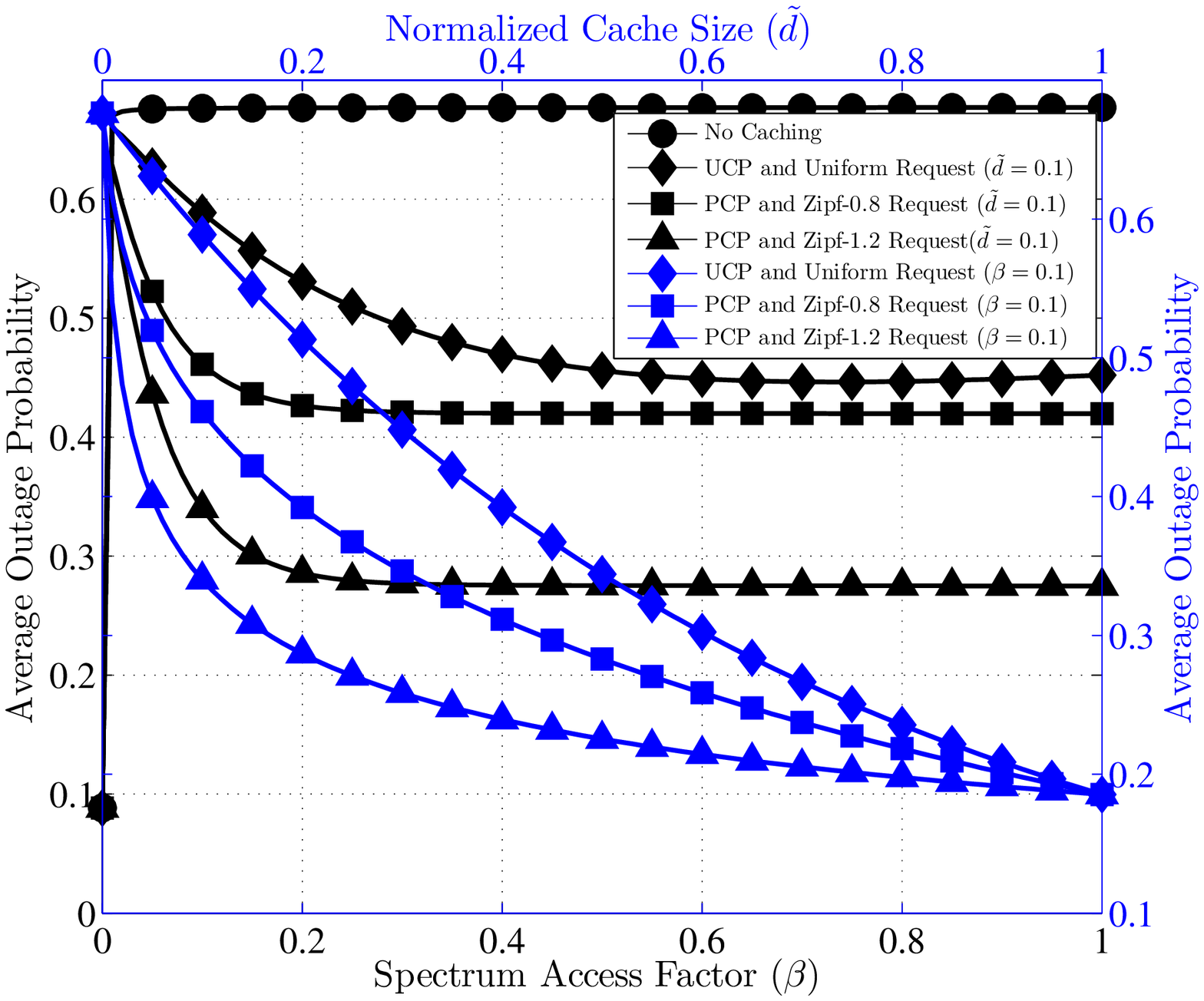}
\vspace{-0.3cm}
\caption{Storage bandwidth tradeoff. $\lambda_{\mathrm{SBS}} = 0.2\mathrm{m}^{-2}$, $\lambda_{\mathrm{MBS}} = 0.0001\mathrm{m}^{-2}$, $R_{\mathrm{SBS}} = 5$m, $R_{\mathrm{MBS}} = 250$m, $\gamma = -10$dB, $\alpha = 4$.}
\label{fig:Storage_BWTradeoff}
\end{figure}%
\indent Fig. \ref{fig:Storage_BWTradeoff} shows the tradeoff between cache size and bandwidth for different content distributions. In case of UCP, the maximum outage probability is 0.2 for a large cache size. For a decreasing cache size, the outage probability increases under uniform distribution. However, for PCP, the outage probability increases much slower as SBSs cache the most popular contents. Meanwhile, for PCP, the maximum outage probability is 0.46 albeit utilizing the whole spectrum ($\beta=1$) while uniform caching attains the maximum outage probability of 0.5. Therefore, caching smartly requires less amount of spectrum  to achieve the same cache hit probability. Hence, PCP makes use of the network resources more efficiently for a constrained cache size. 
Furthermore, PCP is preferred for a fixed bandwidth as it achieves the minimum outage probability. Finally, Fig. \ref{fig:Threshold_Tradeoff} shows the impact of SIR threshold on outage probability, showing that PCP relaxes the SIR threshold requirement for a given outage probability. Furthermore, spectrum utilization without caching induces much higher outage probability than the joint cache-spectrum allocation.
\vspace{-0.1cm}
\section{Conclusion}
In this letter, we examined a cache-enabled small cell network underlying the macro cellular network, in which SBSs strategically store contents. By considering the distribution of SBSs to be a PPP, we analytically derived the outage probability of serving the requested content by jointly considering spectrum allocation and storage constraints. In addition, we characterized the number of SBSs required to satisfy a given cache hit probability for different content distributions. This letter highlights the key tradeoff between cache size, node density and spectrum, underscoring the fact that larger cache size can be leveraged in conjunction with smaller spectrum access factor.
\vspace{-0.2cm}


%
%


\begin{thebibliography}{1}
\vspace{-0.2cm}

\bibitem{Ref3:SCN_2}
S. Singh, H. S. Dhillon and J. G. Andrews, ``Offloading in Heterogeneous Networks: Modeling, Analysis, and Design Insights,'' \textit{IEEE Trans. Wireless Commun.}, vol. 12, no. 5, pp. 2484-2497, May 2013.

\bibitem{ref:Spec_Utilization1}
P. Luoto et. al, ``Co-Primary Multi-Operator Resource Sharing for Small Cell Networks,'' \textit{IEEE Trans. Wireless Commun.}, vol. 14, no. 6, pp. 3120 - 3130, Feb. 2015.

\bibitem{ref:Spec_Utilization2}
T. Q. S. Quek, G. de la Roche, I. Guvenc, and M. Kountouris, \textit{Small Cell Networks: Deployment, PHY Techniques, and Resource Management}. New York, USA: Cambridge University Press, Sept. 2012.

\bibitem{Ref:Boc_5G}
F. Boccardi, et al., ``Five Disruptive Technology Directions for 5G,'' \textit{IEEE Commun. Mag.}, vol. 52, no.2, pp. 74-80, Feb 2014.

\bibitem{Ref6:Intro1}
E. Bastug, M. Bennis, and M. Debbah, ``Living on the Edge: The Role of Proactive Caching in 5G Wireless Networks,'' \textit{IEEE Commun. Mag.}, vol. 52, no. 8, pp. 82-89, Aug. 2014.

\bibitem{Ref7:lit_review3}
J. Dai, et al., ``Collaborative Caching in Wireless Video Streaming Through Resource Auctions,'' \textit{IEEE J. Sel. Areas Commun.}, vol. 30, no. 2, pp. 458-466, Feb. 2012.

\bibitem{Ref7:lit_review4}
E. Bastug, M. Bennis, M. Kountouris, and M. Debbah, ``Cache-enabled Small Cell Networks: Modeling and Tradeoffs,'' \textit{EURASIP J. Wireless Commun. and Net.}, vol. 2015, no. 1, pp. 41, 2015.

\bibitem{Ref7:lit_review5}
T. Syed, M. Bennis, P. H. J. Nardelli, and M. Latva-aho, ``Modeling and Analysis of Content Caching in Wireless Small Cell Networks,'' \textit{preprint}, 2015, http://arxiv.org/pdf/1507.00182.pdf.

\bibitem{Ref8:Caching_Ref2}
M. Leconte, M. Lelarge, and L. Massoulie, ``Adaptive Replication in Distributed Content Delivery Networks,'' \textit{preprint}, 2014, http://arxiv.org/pdf/1401.1770v1.

\bibitem{Ref8:Caching_Ref4}
P. H. J. Nardelli, P. Cardieri, and M. Latva-Aho, ``Efficiency of Wireless Networks under Different Hopping Strategies,'' \textit{IEEE Trans. Wireless. Commun}, vol. 11, no. 1, pp. 15-20, Apr 2012.

\bibitem{Ref9:Coverage}
J. G. Andrews, F. Baccelli, and R. K. Ganti, ``A Tractable Approach to Coverage and Rate in Cellular Networks,'' \textit{IEEE Trans. on Communications}, vol. 59, no. 11, pp. 3122 – 3134, Nov. 2011.

\vspace{-0.8cm}

\end{thebibliography}
\end{document}